\documentclass[11pt,twoside]{article}
\usepackage{epsfig}
\usepackage{amsfonts}
\usepackage{amssymb}
\usepackage{amstext}
\usepackage{amsmath}
\usepackage{xspace}
\usepackage{shadow}
\usepackage{a4wide}
\usepackage{anysize}
\usepackage{latexsym}
\usepackage{amsmath}
\usepackage{amssymb}
\usepackage{epsfig}
\usepackage{tabls}
\usepackage[dvips]{color}
\usepackage{fullpage}
\usepackage{times}
\usepackage{tabularx}
\usepackage{xspace}

\newtheorem{theorem}{Theorem}

\newtheorem{corollary}[theorem]{Corollary}
\newtheorem{lemma}[theorem]{Lemma}
\newtheorem{cl}[theorem]{Claim}
\newcommand\ignore[1]{}

\newenvironment{pf}{\noindent{\bf Proof:  }}{\hfill\rule{2mm}{2mm}\medskip}
\newenvironment{pfof}[1]{\noindent{\bf Proof of #1:  }}{\hfill\rule{2mm}{2mm}\medskip}

\def\sse{\subseteq}
\def\opt{{\sf Opt}}
\def\alg{{\sf ALG}}
\def\m{\mathcal{M}}
\def\is{\mathcal{I}}
\def\bs{\mathcal{B}}
\def\gs{\mathcal{G}}
\def\us{\mathcal{U}}
\def\dom{\mathcal{D}}
\def\squareforqed{\hbox{\rlap{$\sqcap$}$\sqcup$}}
\def\qed{\ifmmode\else\unskip\quad\fi\squareforqed}
\def\smartqed{\def\qed{\ifmmode\squareforqed\else{\unskip\nobreak\hfil
\penalty50\hskip1em\null\nobreak\hfil\squareforqed
\parfillskip=0pt\finalhyphendemerits=0\endgraf}\fi}}

\title{Non-monotone Submodular Maximization under\\ Matroid and Knapsack Constraints}

\author{
Jon Lee \thanks{IBM T.J. Watson Research Center, {\tt jonlee@us.ibm.com}}
\and
Vahab S. Mirrokni \thanks{Google Research, NYC {\tt mirrokni@google.com}}
\and
Viswanath Nagarajan \thanks{Carnegie Mellon University, {\tt viswanat@andrew.cmu.edu}}
\and
Maxim Sviridenko \thanks{IBM T.J. Watson Research Center, {\tt sviri@us.ibm.com}}
 }

\begin{document}

\begin{titlepage}
\maketitle

\begin{abstract} Submodular function maximization is a central problem in
combinatorial optimization, generalizing many important problems including Max Cut in directed/undirected graphs and in hypergraphs, certain constraint satisfaction problems, maximum entropy
sampling, and maximum facility location problems. Unlike submodular minimization, submodular maximization is NP-hard. \ignore{For the special case of monotone submodular functions, tight
$\left(1-{1\over e}\right)$-approximation algorithms have been developed for the submodular maximization problem subject to a matroid or multiple knapsack constraints~\cite{NWF78,F98,V08,KST09}.
For the problem of maximizing a non-monotone submodular function, Feige, Mirrokni, and Vondr\'ak recently developed a $2\over 5$-approximation algorithm~\cite{FMV07}, however, their algorithms
do not handle side constraints.} In this paper, we give the first constant-factor approximation algorithm for maximizing any non-negative submodular function subject to multiple matroid or
knapsack constraints. We emphasize that our results are for {\em non-monotone} submodular functions. In particular, for any constant $k$, we present a $\left({1\over k+2+{1\over
k}+\epsilon}\right)$-approximation for the submodular maximization problem under $k$ matroid constraints, and a $\left({1\over 5}-\epsilon\right)$-approximation algorithm for this problem
subject to $k$ knapsack constraints ($\epsilon>0$ is any constant). We improve the approximation guarantee of our algorithm to ${1\over k+1+{1\over k-1}+\epsilon}$ for $k\ge 2$ partition matroid
constraints. This idea also gives a $\left({1\over k+\epsilon}\right)$-approximation for maximizing a {\em monotone} submodular function subject to $k\ge 2$ partition matroids, which improves
over the previously best known guarantee of $\frac{1}{k+1}$.

%Our main technique for the above results is a new local search algorithm.
%In the design of these algorithms, we also use structural properties of matroids,
%a fractional relaxation of submodular functions, and a randomized rounding technique.
\end{abstract}

\thispagestyle{empty} \setcounter{page}{0}

\end{titlepage}

\section{Introduction}
In this paper, we consider the problem of maximizing a {\em nonnegative submodular function $f$, defined on a ground set $V$, subject to matroid constraints or knapsack constraints}. A function
$f: 2^V \rightarrow {\mathbb R}$ is \emph{submodular} if for all $S, T \subseteq V$, $f(S \cup T) + f(S \cap T) \leq f(S) + f(T)$. \ignore{An alternative definition of submodularity is the
property of \emph{decreasing marginal values}: For all $A \subseteq B \subseteq V$ and $x \in V \setminus B$,
\[
f(B \cup \{x\}) - f(B) \leq f(A \cup \{x\}) - f(A).
\]
}Throughout, we assume that our submodular function $f$ is given by a \emph{value oracle}; i.e., for a given set $S\subseteq V$, an algorithm can query an oracle to find its value $f(S)$.
Furthermore, all submodular functions we deal with are assumed to be non-negative. We also denote the ground set $V=[n]=\{1,2,\cdots,n\}$.

We emphasize that our focus is on submodular functions that are \emph{not required to be monotone} (i.e., we do \emph{not} require that $f(X)\le f(Y)$ for $X\subseteq Y\subseteq V$).
Non-monotone submodular functions appear in several places including cut functions in weighted directed or undirected graphs or even hypergraphs, maximum facility location, maximum entropy
sampling, and certain constraint satisfaction problems.

Given a weight vector $w$ for the ground set $V$, and a knapsack of capacity $C$, the associated \emph{knapsack constraint} is that the sum of weights of elements in the solution $S$ should not
exceed the capacity $C$, i.e, $\sum_{j\in S} w_j \le C$. In our usage, we consider $k$ knapsack constraints defined by weight vectors $w^i$ and capacities $C_i$, for $i=1,\ldots,k$.

We assume some familiarity with matroids \cite{OxleyBook} and associated algorithmics \cite{SchrijverBook}. Briefly, for a matroid ${\cal M}$, we denote the ground set of ${\cal M}$ by ${\cal
E}({\cal M})$, its set of independent sets by ${\cal I}({\cal M})$, and its set of bases by ${\cal B}({\cal M})$. For a given matroid $\cal M$, the associated \emph{matroid constraint} is $S\in
{\cal I}({\cal M})$ and the associated \emph{matroid base constraint} is $S\in {\cal B}({\cal M})$. \ignore{As is standard, ${\cal M}$ is a \emph{uniform matroid} of rank $r$ if ${\cal I}({\cal
M}):=\{X\subseteq {\cal E}({\cal M}) ~:~ |X|\le r\}$, and a \emph{partition matroid} is the direct sum of uniform matroids. Note that uniform matroid constraints are equivalent to cardinality
constraints, i.e, $\vert S\vert \le k$.} In our usage, we deal with $k$ matroids ${\cal M}_1,\ldots,{\cal M}_k$ on the common ground set $V:={\cal E}({\cal M}_1)=\cdots={\cal E}({\cal M}_k)$
(which is also the ground set of our submodular function $f$), and we let ${\cal I}_i:={\cal I}({\cal M}_i)$ for $i=1,\ldots,k$.\\

%In what follows, we consider the problem of
%\emph{maximizing a nonnegative submodular function subject to  $k$ knapsack constraints or $k$ matroid constraints}.
%This means, given a submodular function $f: 2^X \rightarrow \RR_+$ and extra matroid or knapsack constraints,
%we want to find a subset
%$S \subseteq X$ such that $S$ satisfies the extra constraints and $f(S)$ is maximized.

%\paragraph{Background}
{\bf \noindent Background.} Optimizing submodular functions is a central subject in operations research and combinatorial optimization~\cite{L83}. This problem  appears in many important
optimization problems including cuts in graphs~\cite{GW95,Q95,FFI00}, rank function of matroids~\cite{E70,F97}, set covering problems~\cite{F98}, plant location
problems~\cite{CFN77a,CFN77b,CNW90,AS99}, and certain satisfiability problems~\cite{Hastad,FG95}, and maximum entropy sampling~\cite{ME1,ME2}. Other than many heuristics that have been developed
for optimizing these functions~\cite{GSTT99,GTT99,K89,RS89,LNW96}, many exact and constant-factor approximation algorithms are also known for this
problem~\cite{NWF78,NWF78II,Lex00,FFI00,FMV07,SF,GHIM}. In some settings such as set covering or matroid optimization, the relevant submodular functions are monotone.
%, meaning that $f(S) \leq f(T)$ whenever $S \subseteq T$.
Here, we are more interested in the general case where $f(S)$ is not necessarily monotone.
%Cuts in undirected graphs and hypergraphs yield
%\emph{symmetric} submodular functions, satisfying $f(S) = f(\bar{S})$
%for all sets $S$. Symmetric submodular functions have been considered widely
%in the literature~\cite{F83,Q95}. It appears that symmetry allows better
%approximation results, and thus deserves separate attention.

Unlike submodular minimization~\cite{Lex00,FFI00}, submodular function maximization is NP-hard as it generalizes many NP-hard problems, like Max-Cut~\cite{GW95,FG95} and maximum facility
location~\cite{CFN77a,CFN77b,AS99}. Other than generalizing combinatorial optimization problems like Max Cut~\cite{GW95}, Max Directed Cut~\cite{Alimonti97,HZ01}, hypergraph cut problems,
maximum facility location~\cite{AS99,CFN77a,CFN77b}, and certain restricted satisfiability problems~\cite{Hastad,FG95}, maximizing non-monotone submodular functions have applications in a
variety of problems, e.g, computing the core value of supermodular games~\cite{SU07}, and optimal marketing for revenue maximization over social networks~\cite{HMS08}. As an example, we describe
one important application in the statistical design of experiments. The \emph{maximum entropy sampling problem} is as follows: Let $A$ be the $n$-by-$n$ covariance matrix of a set of Gaussian
random variables indexed by $[n]$. For $S\subseteq [n]$, let $A[S]$ denote the principal submatrix of $A$ indexed by $S$. It is well known that (up to constants depending on $|S|$), $\log\det
A[S]$ is the entropy of the random variables indexed by $S$. Furthermore, $\log\det A[S]$ is submodular on $[n]$. In applications of locating environmental monitoring stations, it is desired to
choose $s$ locations from $[n]$ so as to maximize the entropy of the associated random variables, so that problem is precisely one of maximizing a non-monotone submodular function subject to a
cardinality constraint. Of course a cardinality constraint is just a matroid base constraint for a uniform matroid. We note that the entropy function is not even approximately monotone (see
\cite{Krause}). The maximum entropy sampling problem has mostly been studied from a computational point of view, focusing on calculating optimal solutions for moderate-sized instances (say
$n<200$) using mathematical programming methodologies (e.g, see \cite{ME1,ME2,ME3,ME4,ME5,ME6}), and our results provide the first set of algorithms with provable constant-factor approximation
guarantee.

%For most of the well-studied special cases of this problem, a better than $1\over 2$-approximation can be achieved using
%semidefinite programming: $0.878$ for Max Cut \cite{GW95}, $0.859$ for Max Directed Cut~\cite{FG95}, and $0.828$ for
%maximum facility location problems~\cite{AS99}.
Recently, a $2\over 5$-approximation was developed for maximizing non-negative non-monotone submodular functions without any side constraints~\cite{FMV07}. This algorithm also provides a tight
$1\over 2$-approximation algorithm for maximizing a symmetric\footnote{The function
 $f: 2^V \rightarrow {\mathbb R}$  is symmetric if for all  $S\subseteq V$, $f(S) = f(V\setminus S)$.
For example,  cut functions in undirected graphs are well-known
examples of symmetric (non-monotone) submodular functions}
submodular function~\cite{FMV07}. However, the algorithms
developed  in \cite{FMV07} for non-monotone submodular
maximization do not handle any extra constraints.

For the problem of maximizing a monotone submodular function subject to a matroid or multiple knapsack constraints, tight $\left(1-{1\over e}\right)$-approximation are
known~\cite{NWF78,CCPV07,V08,Sviri,KST09}. Maximizing monotone submodular functions over $k$ matroid constraints was considered in~\cite{NWF78II}, where a
$\left(\frac{1}{k+1}\right)$-approximation was obtained. This bound is currently the best known ratio, even in the special case of partition matroid constraints. However, none of these results
generalize to non-monotone submodular functions.

Better results are known either for specific submodular functions or for special classes of matroids. A $\frac{1}{k}$-approximation algorithm using local search was designed in  \cite{RS} for
the problem of maximizing a linear function subject to $k$ matroid constraints. Constant factor approximation algorithms are known for the problem of maximizing directed cut \cite{AHS} or
hypergraph cut \cite{AS} subject to a uniform matroid (i.e. cardinality) constraint.

Hardness of approximation results are known even for the special case of maximizing a linear function subject to $k$ partition matroid constraints. The best known lower bound is an
$\Omega(\frac{k}{\log k})$ hardness of approximation~\cite{HSS}. Moreover, for the unconstrained maximization of non-monotone submodular functions, it has been shown that achieving a factor
better than $1\over 2$ cannot be done using a subexponential number of value queries~\cite{FMV07}.\\

{\bf \noindent Our Results.} In this paper, we give the first constant-factor approximation algorithms for maximizing a non-monotone submodular function subject to multiple matroid constraints,
or multiple knapsack constraints. More specifically, we give the following new results (below $\epsilon>0$ is any constant).

\noindent {\bf (1)} For every constant $k\ge 1$, we present a $\left({1\over k+2+{1\over k}+\epsilon}\right)$-approximation algorithm for maximizing any non-negative submodular function subject
to $k$ matroid constraints (Section~\ref{sec:k-mat}). This implies a $\left({1\over 4+\epsilon}\right)$-approximation algorithm for maximizing non-monotone submodular functions subject to a
single matroid constraint. Moreover, this algorithm is a $\left({1\over k+2+\epsilon}\right)$-approximation in the case of \emph{symmetric} submodular functions. Asymptotically, this result is
nearly best possible because there is an $\Omega(\frac{k}{\log k})$ hardness of approximation, even in the monotone case~\cite{HSS}.
%For the special case of one matroid constraint (or cardinality constraint),
%there is a gap between the approximation factor of our algorithm, $1\over 4$, and the hardness factor
%of $1\over 2$~\cite{FMV07}.

\noindent {\bf (2)} For every constant $k\ge 1$, we present a $\left({1\over 5}-\epsilon\right)$-approximation algorithm for maximizing any nonnegative submodular function subject to a
$k$-dimensional knapsack constraint (Section~\ref{sec:k-knap}). To achieve the approximation guarantee, we first give a $\left({1\over 4} - \epsilon\right)$-approximation algorithm for a
fractional relaxation (similar to the fractional relaxation used in \cite{V08}). We then use a simple randomized rounding technique to convert a fractional solution to an integral one. A similar
method was recently used in \cite{KST09} for maximizing a monotone submodular function over knapsack constraints, but neither their algorithm for the fractional relaxation, nor their  rounding
method is  directly applicable to non-monotone submodular functions.
%Our main
%contribution here is a ${1\over 4} - \epsilon$-approximation
%algorithm for the fractional relaxation.

\noindent {\bf (3)} For submodular maximization under $k\ge 2$ {\em partition matroid} constraints, we obtain improved approximation guarantees (Section~\ref{sec:k-partn}). We  give a
$\left({1\over k+1+{1\over k-1}+\epsilon}\right)$-approximation algorithm for maximizing non-monotone submodular functions subject to $k$ partition matroids. \ignore{This also implies a
$\left({1\over 4}-\epsilon\right)$-approximation algorithm for maximizing a nonnegative submodular function subject to a (non-bipartite) matching constraint.} Moreover, our idea gives a
$\left({1\over k+\epsilon}\right)$-approximation algorithm for maximizing a monotone submodular function subject to $k\ge 2$ partition matroid constraints.  This is an improvement over the
previously best known bound of ${1\over k+1}$ from \cite{NWF78II}.

\noindent {\bf (4)} Finally, we study submodular maximization subject to a matroid {\em base} constraint in Appendix~\ref{sec:mat-base}. We give a $\left(\frac13-\epsilon\right)$-approximation
in the case of symmetric submodular functions. Our result for general submodular functions only holds for special matroids: we obtain a $(\frac16 -\epsilon)$-approximation when the matroid
contains two disjoint bases. In particular, this implies a $\left({1\over 6}-\epsilon\right)$-approximation for the problem of maximizing any non-negative submodular function subject to an exact
cardinality constraint. Previously, only special cases of directed cut \cite{AHS} or hypergraph cut \cite{AS} subject to an exact cardinality constraint were considered.

Our main technique for the above results is local search. Our local search algorithms are different from the previously used variant of local search for unconstrained maximization of a
non-negative submodular function~\cite{FMV07}, or the local search algorithms used for Max Directed Cut~\cite{Alimonti97,HZ01}. In the design of our algorithms, we also use structural properties
of matroids, a fractional relaxation of submodular functions, and a randomized rounding technique.

%$$
%\begin{array}{|| c || c | c | c | c | c ||} \hline
% \mbox{Problem/Constraint} & \mbox{Mon/} & \mbox{Non-Adapt.} & \mbox{Det.Adapt.} & \mbox{Rnd.Adapt.}
%    & \mbox{Hardness} \\
%\hline \hline
%\mbox{Approximation Algorithm} & 1/2 & 1/2 & 1/2 & 1/2 & 5/6 \\
%\hline
%\mbox{Inapproximability} & 1/4 & 1/3 & 1/3 & 2/5 & 3/4 \\
%\hline
% \end{array} $$

\section{Matroid Constraints\label{sec:k-mat}}
In this section, we give an approximation algorithm for submodular maximization subject to  $k$ matroid constraints. The problem is as follows: Let $f$ be a \emph{non-negative} submodular
function defined on ground set $V$. Let $\m_1,\cdots,\m_k$ be $k$ arbitrary matroids on the common ground set $V$. For each matroid $\m_j$ (with $j\in[k]$) we denote the set of its independent
sets by $\is_j$. We consider the following problem:
\begin{equation}\label{prob:k-mat}
\max \left\{f(S) ~:~ S\in \cap_{j=1}^k {\cal I}_j  \right\}.
\end{equation}

We give an approximation algorithm for this problem using value queries to $f$ that runs in time $n^{O(k)}$. The starting point  is the following local search algorithm. Starting with
$S=\emptyset$, repeatedly perform one of the following local improvements:
\begin{itemize}
 \item {\bf Delete operation.} If $e\in S$ such that $f(S\setminus \{e\})>f(S)$, then $S\leftarrow
 S\setminus\{e\}$.
 \item  {\bf Exchange operation.} If $d\in V\setminus S$ and $e_i\in S\cup\{\phi\}$
 (for $1\le i\le k$) are such that $(S\setminus
 \{e_i\})\cup \{d\}\in{\cal I}_i$ for all $i\in[k]$ and $f((S\setminus
 \{e_1,\cdots,e_k\})\cup \{d\})>f(S)$, then $S\leftarrow (S\setminus
 \{e_1,\cdots,e_k\})\cup \{d\}$.
\end{itemize}
\smallskip

When dealing with a single matroid constraint ($k=1$), the local operations correspond to: \emph{delete} an element, \emph{add} an element (i.e. an exchange when no element is dropped),
\emph{swap} a pair of elements (i.e. an element from outside the current set is exchanged with an element from the set). With $k\ge 2$ matroid constraints, we permit more general exchange
operations, involving adding one element and dropping {\em up to} $k$ elements.

Note that the size of any local neighborhood is at most $n^{k+1}$, which implies that each local step can be performed in polynomial time for a constant $k$. Let $S$ denote a locally optimal
solution. Next we prove a key lemma for this local search algorithm, which is used in analyzing our algorithm. Before presenting the lemma, we state a useful exchange property of matroids (see
\cite{SchrijverBook}). Intuitively, this property states that for any two independent sets $I$ and $J$, we can add any element of $J$ to the set $I$, and kick out at most one element from $I$
while keeping the set independent. Moreover, each element of $I$ is allowed to be kicked out by at most one element of $J$. For completeness, a proof is given in Appendix~\ref{app:k-mat}.
\begin{theorem}\label{th:mat-exch}
Let $\m$ be a matroid and $I,J\in {\cal I}({\cal M})$  be two independent sets. Then
there is a mapping $\pi:J\setminus I\rightarrow (I\setminus J)\cup
\{\phi\}$ such that:
\begin{enumerate}
 \item $(I\setminus \pi(b))\cup \{b\}\in {\cal I}({\cal M})$ for all $b\in J\setminus I$.
 \item $|\pi^{-1}(e)|\le 1$ for all $e\in I\setminus J$.
\end{enumerate}
\end{theorem}

\begin{lemma}\label{lem:k-mat-main}
For a local optimal solution $S$ and any $C\in\cap_{j=1}^k \is_j$, $(k+1)\cdot f(S)\ge f(S\cup C) + k\cdot f(S\cap C)$. Additionally for $k=1$, if $S\in\is_1$ is any locally optimal solution
under only the swap operation, and $C\in \is_1$ with $|S|=|C|$, then $2\cdot f(S)\ge f(S\cup C) + f(S\cap C)$.
\end{lemma}
\begin{pf}
The following proof is due to Jan Vondr\'ak~\cite{v-pc}. Our original proof~\cite{lmns} was more complicated-- we thank Jan for letting us present this simplified proof.

For each matroid $\m_j$ ($j\in[k]$), because both $C,S\in\is_j$ are independent sets, Theorem~\ref{th:mat-exch} implies a mapping $\pi_j:C\setminus S\rightarrow (S\setminus C)\cup \{\phi\}$ such
that:
\begin{enumerate}
 \item $(S\setminus \pi_j(b))\cup \{b\}\in\is_j$ for all $b\in C\setminus S$.
 \item $|\pi_j^{-1}(e)|\le 1$ for all $e\in S\setminus C$.
\end{enumerate}
When $k=1$ and $|S|=|C|$, Corollary 39.12a from \cite{SchrijverBook} implies the stronger condition that $\pi_1:C\setminus S\rightarrow S\setminus C$ is in fact a \emph{bijection}.

For each $b\in C\setminus S$, let $A_b=\{\pi_1(b),\cdots \pi_k(b)\}$. Note that $(S\setminus A_b)\cup \{b\}\in \cap_{j=1}^k \is_j$ for all $b\in C\setminus S$. Hence $(S\setminus A_b)\cup \{b\}$
is in the local neighborhood of $S$, and by local optimality under exchanges:
\begin{equation}\label{eq:local1} f(S)\ge
f\left(\left(S\setminus A_b\right)\cup\{b\}\right), \qquad \qquad \forall b\in C\setminus S.
\end{equation}
In the case $k=1$ with $|S|=|C|$, these are only \emph{swap} operations (because $\pi_1$ is a bijection here).

By the property of mappings $\{\pi_j\}_{j=1}^k$, each element $i\in S\setminus C$ is contained in $n_i\le k$ of the sets $\{A_b\mid b\in C\setminus S\}$; and elements of $S\cap C$ are contained
in none of these sets. So the following inequalities are implied by local optimality of $S$ under deletions.
\begin{equation}\label{eq:local2}(k-n_i)\cdot f(S)\ge (k-n_i)\cdot f(S\setminus
\{i\}),\qquad \qquad \forall i\in S\setminus C.\end{equation} Note that these inequalities are not required when $k=1$ and $|S|=|C|$ (then $n_i=k$ for all $i\in S\setminus C$).

For any $b\in C\setminus S$, we have (below, the first inequality is submodularity and the second is from~\eqref{eq:local1}):
\begin{equation*}
f(S\cup \{b\}) - f(S)\le f\left( \left(S\setminus A_b\right) \cup \{b\}\right) - f\left(S\setminus A_b\right) \le f(S) - f\left(S\setminus A_b\right)
\end{equation*}
Adding this inequality over all $b\in C\setminus S$ and using submodularity,
\begin{equation*}
f(S\cup C)-f(S)\le \sum_{b\in C\setminus S}\left[ f(S\cup \{b\}) - f(S) \right] \le \sum_{b\in C\setminus S}\left[ f(S) - f\left(S\setminus A_b\right) \right]
\end{equation*}
Adding to this, the inequalities~\eqref{eq:local2}, i.e. $0\le (k-n_i)\cdot \left[ f(S) - f(S\setminus \{i\}) \right]$ for all $i\in S\setminus C$,
\begin{eqnarray}
f(S\cup C)-f(S)&\le &\sum_{b\in C\setminus S}\left[ f(S) - f\left(S\setminus A_b\right) \right] + \sum_{i\in S\setminus C} (k-n_i)\cdot \left[ f(S) - f(S\setminus \{i\}) \right] \notag\\
&=& \sum_{l=1}^\lambda \left[ f(S) - f\left(S\setminus T_l\right) \right] \label{eq:exch-lem}
\end{eqnarray}
where  $\lambda=|C\setminus S| + \sum_{i\in S\setminus C} (k-n_i)$ and $\{T_l\}_{l=1}^\lambda$ is some collection of subsets of $S\setminus C$ such that each $i\in S\setminus C$ appears in {\em
exactly} $k$ of these subsets. Let $s=|S|$ and $|S\cap C|=c$; number the elements of $S$ as $\{1,2,\cdots,s\}=[s]$ such that $S\cap C = \{1,2,\cdots,c\}=[c]$. Then for any $T\sse S\setminus C$,
by submodularity: $f(S)-f(S\setminus T)\le \sum_{p\in T} \left[ f([p]) - f([p-1])\right]$. Using this in~\eqref{eq:exch-lem}, we obtain:
\begin{equation*}
f(S\cup C)-f(S) \le \sum_{l=1}^\lambda \sum_{p\in T_l} \left[ f([p]) - f([p-1])\right] = k\sum_{i=c+1}^s \left[ f([i]) - f([i-1])\right] = k\cdot \left( f(S) - f(S\cap C)\right)
\end{equation*}
The second last equality follows from $S\setminus C = \{c+1,\cdots,s\}$ and the fact that each element of $S\setminus C$ appears in exactly $k$ of the sets $\{T_l\}_{l=1}^\lambda$. The last
equality is due to a telescoping summation. Thus we obtain $(k+1)\cdot f(S)\ge f(S\cup C) + k\cdot f(S\cap C)$, giving the claim.

Observe that when $k=1$ and $|S|=|C|$, we only used the inequalities~\eqref{eq:local1} from the local search, which are only swap operations. Hence in this case, the statement also holds for any
solution $S$ that is locally optimal under only swap operations. In the general case, we use both inequalities~\eqref{eq:local1} (from exchange operations) and inequalities~\eqref{eq:local2}
(from deletion operations).
\end{pf}

A simple consequence of Lemma \ref{lem:k-mat-main}  implies bounds analogous to known approximation factors \cite{NWF78II,RS} in the cases when the submodular function $f$ has additional
structure.
\begin{corollary}
For a locally optimal solution $S$ and any $C\in\cap_{j=1}^k
\is_j$ the following inequalities hold:
\begin{enumerate}
\item $f(S)\ge f(C)/(k+1)$ if function $f$ is monotone, \item
$f(S)\ge f(C)/k$ if function $f$ is linear.
\end{enumerate}
\end{corollary}

The local search algorithm defined above could run for an exponential amount of time until it reaches a locally optimal solution. The standard approach is to consider an approximate local
search. In Appendix~\ref{app:k-mat}, we show an inequality (Lemma~\ref{lem:apx-local}) analogous to Lemma \ref{lem:k-mat-main} for approximate local optimum.

\begin{figure}
\begin{center}
\fbox{\parbox{6in}{
\begin{small}
\begin{description}
\item{\bf Approximate Local Search Procedure $B$:} $\ $

{\bf Input: } Ground set $X$ of elements and value oracle access
to submodular function $f$.

\begin{enumerate}
 \item Let $\{v\}$ be a singleton set with the maximum value $f(\{v\})$ and let $S=\{v\}$.
 \item While there exists  the following delete or exchange local operation that increases the value of $f(S)$ by a factor of at least
 $1 + {\epsilon\over n^4}$, then apply the local operation and update $S$ accordingly.
 \begin{itemize}
    \item {\bf Delete operation on $S$.} If $e\in S$ such that $f(S\setminus \{e\})\ge (1+{\epsilon\over n^4}) f(S)$, then $S\leftarrow
   S\setminus\{e\}$.
   \item  {\bf Exchange operation on $S$.} If $d\in X\setminus S$ and $e_i\in S\cup\{\phi\}$ (for $1\le i\le k$) are such that $(S\setminus
   \{e_i\})\cup \{d\}\in\is_i$ for all $i\in[k]$ and $f((S\setminus
   \{e_1,\cdots,e_k\})\cup \{d\})>(1+{\epsilon\over n^4}) f(S)$, then $S\leftarrow (S\setminus
   \{e_1,\cdots,e_k\})\cup \{d\}$.
   \end{itemize}
\end{enumerate}
\end{description}
\end{small}
}}
\end{center}
 \caption{\label{fig:algorithmb}The approximate local search procedure.}
\end{figure}

\begin{figure}
\begin{center}
\fbox{\parbox{6in}{
\begin{small}
\begin{description}
\item{\bf Algorithm $A$:} $\ $

\begin{enumerate}
 \item Set $V_1=V$.
 \item For $i=1,\cdots,k+1$, do:
 \begin{enumerate}
  \item Apply the approximate local search procedure $B$ on the ground set $V_i$ to obtain a solution $S_i\sse V_i$ corresponding to the problem:
  \begin{equation}\label{eq:iter-prob}\max\{f(S) ~:~ S\in\cap_{j=1}^k \is_j,~S\sse
  V_i\}.\end{equation}
  \item Set $V_{i+1}=V_i\setminus S_i$.
 \end{enumerate}
 \item Return the solution corresponding to $\max\{f(S_1),\cdots,f(S_{k+1})\}$.
\end{enumerate}
\end{description}
\end{small}
}}
\end{center}
\caption{\label{fig:algorithma} Approximation algorithm for
submodular maximization under $k$ matroid constraints.}
\end{figure}

\begin{theorem}\label{th:main}
Algorithm $A$ in Figure \ref{fig:algorithma} is a $\left({1\over
(1+\epsilon)(k+2+\frac{1}{k})}\right)$-approximation algorithm for
maximizing a non-negative submodular function subject to any $k$
matroid constraints, running in time $n^{O(k)}$.
%Moreover, for symmetric submodular functions,
%the approximation factor of this algorithm is $({1\over k+2}-o(1))$
\end{theorem}
\begin{pf}
Bounding the running time of Algorithm $A$ is easy and we leave it to Appendix~\ref{app:k-mat}. Here, we prove the performance guarantee of Algorithm $A$. Let $C$ denote the optimal solution to
the original problem $\max\{f(S) ~:~ S\in\cap_{j=1}^k \is_j,~S\sse V\}$. Let $C_i=C\cap V_i$ for each $i\in[k+1]$; so $C_1=C$. Observe that $C_i$ is a feasible solution to the
problem~\eqref{eq:iter-prob} solved in the $i$th iteration. Applying Lemma~\ref{lem:apx-local} to problem~\eqref{eq:iter-prob} using the local optimum $S_i$ and solution $C_i$, we obtain:
\begin{equation}\label{eq:3}(1+\epsilon)(k+1)\cdot f(S_i)\ge
f(S_i\cup C_i)+k\cdot f(S_i\cap C_i)\qquad \forall 1\le i\le k+1,\end{equation} Using $f(S)\ge \max_{i=1}^{k+1} f(S_i)$, we add these $k+1$ inequalities and simplify inductively as follows.
\begin{cl}\label{cl:ind-ineq}
For any $1\le l\le k+1$, we have:
\begin{eqnarray*}
(1+\epsilon)(k+1)^2\cdot f(S) &\ge &(l-1)\cdot f(C) + f(\cup_{p=1}^{l} S_p \cup C_1)+\sum_{i=l+1}^{k+1} f(S_i\cup C_i) \\
&&+ \sum_{p=1}^{l-1} (k-l+p) \cdot f(S_p\cap C_p) +k\cdot \sum_{i=l}^{k+1} f(S_i\cap C_i).
\end{eqnarray*}
\end{cl}
\begin{pf}
We argue by induction on $l$. The base case $l=1$ is trivial, by just considering the sum of the $k+1$ inequalities in statement~\eqref{eq:3} above. Assuming the statement for some value $1\le
l< k+1$, we prove the corresponding statement for $l+1$.
\begin{eqnarray*}
(1+\epsilon)(k+1)^2\cdot f(S)&\ge& (l-1)\cdot f(C) + f(\cup_{p=1}^{l} S_p \cup C_1)\\
&&+\sum_{i=l+1}^{k+1} f(S_i\cup C_i) + \sum_{p=1}^{l-1} (k-l+p)
f(S_p\cap C_p) +k\cdot \sum_{i=l}^{k+1} f(S_i\cap C_i)\\
&=&(l-1)\cdot f(C) + f(\cup_{p=1}^{l} S_p \cup C_1) + f(S_{l+1}\cup C_{l+1})\\
&&+\sum_{i=l+2}^{k+1} f(S_i\cup C_i) + \sum_{p=1}^{l-1} (k-l+p)
f(S_p\cap C_p) +k\cdot \sum_{i=l}^{k+1} f(S_i\cap C_i)\\
&\ge&(l-1)\cdot f(C) +f(\cup_{p=1}^{l+1} S_p \cup C_1) +
f(C_{l+1})\\
&&+\sum_{i=l+2}^{k+1} f(S_i\cup C_i) + \sum_{p=1}^{l-1} (k-l+p)
f(S_p\cap C_p) +k\cdot \sum_{i=l}^{k+1} f(S_i\cap C_i)\\
&=&(l-1)\cdot f(C) +f(\cup_{p=1}^{l+1} S_p \cup C_1) +f(C_{l+1}) +\sum_{p=1}^l f(S_p\cap C_p)\\
&&+\sum_{i=l+2}^{k+1} f(S_i\cup C_i) + \sum_{p=1}^{l} (k-l-1+p)
f(S_p\cap C_p) +k\cdot \sum_{i=l+1}^{k+1} f(S_i\cap C_i)\\
&\ge& l\cdot f(C) +f(\cup_{p=1}^{l+1} S_p \cup C_1)\\
&&+\sum_{i=l+2}^{k+1} f(S_i\cup C_i) + \sum_{p=1}^{l} (k-l-1+p)
f(S_p\cap C_p) +k\cdot \sum_{i=l+1}^{k+1} f(S_i\cap C_i).
\end{eqnarray*}

The first inequality is the induction hypothesis, and the next two inequalities follow from submodularity using $\left(\cup_{p=1}^l S_p\cap C_p\right)\cup C_{l+1}=C$.
\end{pf}

Using the statement of Claim~\ref{cl:ind-ineq} when $l=k+1$, we obtain $(1+\epsilon)(k+1)^2\cdot f(S)\ge k\cdot f(C)$.
\end{pf}

Finally, we give an improved approximation algorithm for symmetric submodular functions $f$, that satisfy $f(S) = f(\overline{S})$ for all $S\subset V$. Symmetric submodular functions have been
considered widely in the literature~\cite{F83,Q95}, and it appears that symmetry allows for better approximation results and thus deserves separate attention.

\begin{theorem}
There is a $\left({1\over (1+\epsilon)(k+2)}\right)$-approximation algorithm for maximizing a non-negative  symmetric submodular functions subject to $k$ matroid constraints.
\end{theorem}
\begin{pf}
The algorithm for symmetric submodular functions is simpler. In this case, we only need to perform \emph{one} iteration of the approximate local search procedure $B$ (as opposed to $k+1$ in
Theorem~\ref{th:main}). Let $C$ denote the optimal solution, and $S_1$ the result of the local search (on $V$). Then Lemma~\ref{lem:k-mat-main} implies:
$$(1+\epsilon)(k+1)\cdot f(S_1)\ge f(S_1\cup C)+k\cdot f(S_1\cap C)\ge f(S_1\cup C)+f(S_1\cap C).$$
Because $f$ is symmetric, we also have $f(S_1)=f(\overline{S_1})$.
Adding these two,
$$(1+\epsilon)(k+2)\cdot f(S_1)\ge f(\overline{S_1}) +f(S_1\cup
C)+f(S_1\cap C)\ge f(C\setminus S_1) + f(S_1\cap C)\ge f(C)~.$$ Thus we have the desired approximation guarantee.\end{pf}

\section{Knapsack constraints}\label{sec:k-knap}
%In this section, we give an approximation algorithm for submodular maximization subject to multiple knapsack constraints. 
Let $f:2^V\rightarrow \mathbb{R}_+$ be a submodular function, and
$w^1,\cdots,w^k$ be $k$ weight-vectors corresponding to knapsacks having capacities $C_1,\cdots,C_k$ respectively. The problem we consider in this section is:
\begin{equation}\label{prob:k-knap}\max\{f(S) ~:~
\sum_{j\in S} w^i_j \le C_i,~\forall 1\le i\le k, S\subseteq V
\}.\end{equation}

By scaling each knapsack, we assume that $C_i=1$ for all $i\in[k]$. We denote $f_{max}=\max\{f(v) ~:~ v\in V\}$. We assume without loss of generality that for every $i\in V$, the singleton
solution $\{i\}$ is feasible for all the knapsacks (otherwise such elements can be dropped from consideration). To solve the above problem, we first define a fractional relaxation of the
submodular function, and give an approximation algorithm for this fractional relaxation. Then, we show how to design an approximation algorithm for the original integral problem using  the
solution for the fractional relaxation. Let $F:[0,1]^n\rightarrow \mathbb{R}_+$, the \emph{fractional relaxation of $f$}, be the `extension-by-expectation' of $f$, i.e.,
$$F(x)=\sum_{S\sse V} f(S)\cdot \Pi_{i\in S}x_i\cdot \Pi_{j\not\in
S}(1-x_j).$$ Note that $F$ is a multi-linear polynomial in variables
$x_1,\cdots,x_n$, and has continuous derivatives of all orders.
Furthermore, as shown in Vondr\'ak~\cite{V08}, for all $i,j\in V$,
$\frac{\partial^2 }{\partial x_j
\partial x_i}F\le 0$ everywhere on $[0,1]^n$; we refer to this condition as \emph{continuous submodularity}.

\noindent Due to lack of space, many proofs of this section appear in Appendix~\ref{app:k-knap}.

{\bf Extending function $f$ on scaled ground sets.} Let $s_i\in \mathbb{Z}_+$ be arbitrary values for each $i\in V$. Define a new ground-set $U$ that contains $s_i$ `copies' of each element
$i\in V$; so the total number of elements in $U$ is $\sum_{i\in V} s_i$. We will denote any subset $T$ of $U$ as $T=\cup_{i\in V} T_i$ where each $T_i$ consists of all copies of element $i\in V$
from $T$. Now define function $g:2^U\rightarrow \mathbb{R}_+$ as $g(\cup_{i\in V} T_i) = F(\cdots,\frac{|T_i|}{s_i},\cdots)$.

Our goal is to prove the useful lemma that $g$ is submodular. In preparation for that, we first establish a couple of claims. The first claim is standard, but we give a proof for the sake of
completeness.

\begin{cl}\label{cl:cont-prop}
Suppose $l:\dom\rightarrow \mathbb{R}$ has continuous partial derivatives everywhere on convex $\dom\sse \mathbb{R}^n$ with $\frac{\partial l}{\partial x_i}(y)\le 0$ for all $y\in\dom$ and $i\in
V$. Then for any $a_1,a_2\in\dom$ with $a_1\le a_2$ coordinate-wise, we have $l(a_1)\ge l(a_2)$.
\end{cl}

Next, we establish the following property of the fractional relaxation
$F$.
\begin{cl}\label{cl:cont-submod}
For any $a,q,d\in [0,1]^n$ with $q+d\in [0,1]^n$ and $a\le q$
coordinate-wise, we have $F(a+d)-F(a)\ge F(q+d)-F(q)$.
\end{cl}

Using the above claims, we are now ready to state and prove the lemma.

\begin{lemma}\label{lem:scaled-submod}
Set function $g$ is a submodular function on ground set $U$.
\end{lemma}

{\bf Solving the fractional relaxation.} We now argue how to obtain a near-optimal fractional feasible solution for maximizing a non-negative submodular function over $k$ knapsack constraints.
Let $w^1,\cdots,w^k$ denote the weight-vectors in each of the $k$ knapsacks such that all knapsacks have capacity 1. The problem we consider here has additional upper bounds $\{u_i\in
[0,1]\}_{i=1}^n$ on variables:
\begin{equation}\label{prob:frac-k-knap}\max\{F(y) ~:~
w^s\cdot y\le 1~~\forall s\in[k],~~0\le y_i\le u_i~~\forall i\in
V\}.\end{equation}

Denote the region $\us:=\{y ~:~ 0\le y_i\le u_i~\forall i\in V\}$. We define a local search procedure to solve problem~\eqref{prob:frac-k-knap}. \ignore{We will always deal with fractional
solutions that satisfy all knapsacks with equality.} We only consider values for each variable from a discrete set of values in $[0,1]$, namely $\gs=\{p\cdot \zeta ~:~ p\in \mathbb{N},~0\le p\le
\frac1\zeta\}$ where $\zeta=\frac{1}{8n^4}$. Let $\epsilon>0$ be a parameter to be fixed later. At any step with current solution $y\in [0,1]^n$, the following local moves are considered:
\begin{itemize}
 \item Let $A,D\sse [n]$ with $|A|,|D|\le k$. Decrease the variables
 $y(D)$ to any values in $\gs$ and increase variables $y(A)$ to any
 values in $\gs$
 such that the resulting solution $y'$ still satisfies all knapsacks and $y'\in\us$.
 If $F(y')>(1+\epsilon)\cdot F(y)$ then set $y\leftarrow y'$.
\end{itemize}
Note that the size of each local neighborhood is $n^{O(k)}$. Let $a$ be the index corresponding to $\max\{u_i\cdot f(\{i\}) ~:~ i\in V\}$. We start the local search procedure with the solution
$y_0$ having $y_0(a)=u_{a}$ and zero otherwise. Observe that for any $x\in \us^n$, $F(x)\le \sum_{i=1}^n u_i\cdot f(\{i\})\le n\cdot u_{a}\cdot f(\{a\})=n\cdot F(y_0)$. Hence the number of
iterations of local search is $O(\frac1\epsilon \log n)$, and the entire procedure terminates in polynomial time. Let $y$ denote a local optimal solution. We first prove the following based on
the discretization $\gs$.
\begin{cl}\label{cl:knap-local-opt}
Suppose $\alpha,\beta\in [0,1]^n$ are such that each has at most $k$ positive coordinates, $y' :=y-\alpha+\beta\in \us$, and $y'$ satisfies all knapsacks. Then $F(y')\le (1+\epsilon)\cdot
F(y)+\frac1{4n^2} f_{max}$.
\end{cl}

For any $x,y\in{\mathbb R}^n$, we define $x\vee y$ and  $x\wedge y$ by $(x\vee y)_j  := \max(x_j,y_j)$  and $(x\wedge y)_j := \min(x_j,y_j)$  for $j\in [n]$.

\begin{lemma}\label{lem:knap-local-sol}
For local optimal $y\in \us\cap \gs^n$ and any $x\in \us$ satisfying the knapsack constraints, we have $(2+2n\cdot\epsilon)\cdot F(y)\ge F(y\wedge x)+F(y\vee x)-\frac{1}{2n}\cdot
f_{max}$.\end{lemma}
\begin{pf}
For the sake of this proof, we assume that each knapsack $s\in[k]$ has a dummy element (which has no effect on function $f$) of weight 1 in knapsack $s$ (and zero in all other knapsacks), and
upper-bound of 1. So any fractional solution can be augmented to another of the same $F$-value, while satisfying all knapsacks at equality. We augment $y$ and $x$ using dummy elements so that
both satisfy all knapsacks at equality: this does not change any of the values $F(y)$, $F(y\wedge x)$ and $F(y\vee x)$.
%We have $y\wedge x =\langle\min(x_i,y_i)\rangle_{i=1}^n$.
Let $y'=y-(y\wedge x)$ and $x'=x-(y\wedge x)$. Note that for all $s\in [k]$, $w^s\cdot y'=w^s\cdot x'$ and let $c_s=w^s\cdot x'$. We will decompose $y'$ and $x'$ into an equal number of terms as
$y'=\sum_t \alpha_t$ and $x'=\sum_t \beta_t$ such that the $\alpha$s and $\beta$s have small support, and $w^s\cdot \alpha_t=w^s\cdot \beta_t$ for all $t$ and $s\in [k]$.
\begin{small}\begin{enumerate}
 \item Initialize $t\leftarrow 1$, $\gamma\leftarrow 1$, $x''\leftarrow x'$, $y''\leftarrow y'$.
  \item While $\gamma>0$, do:
\begin{enumerate}
 \item \label{step:fknap-1}Consider $LP_x := \{z\ge 0 ~:~ z\cdot w^s=c_s,~\forall s\in[k]\}$
 where the variables are restricted to indices $i\in [n]$ with
 $x''_i>0$. Similarly $LP_y := \{z\ge 0 ~:~ z\cdot w^s=c_s,~\forall s\in[k]\}$
 where the variables are restricted to indices $i\in [n]$ with
 $y''_i>0$. Let $u\in LP_x$ and $v\in LP_y$ be extreme points.
 \item \label{step:fknap-2}Set $\delta_1=\max\{\chi ~:~ \chi \cdot u\le x''\}$
$\delta_2=\max\{\chi ~:~ \chi \cdot v\le y''\}$, and $\delta=\min\{\delta_1,\delta_2\}$.
 \item \label{step:fknap-3}Set $\beta_t\leftarrow \delta\cdot u$, $\alpha_t\leftarrow \delta\cdot
 v$, $\gamma\leftarrow \gamma-\delta$, $x''\leftarrow x''-\beta_t$,
 and $y''\leftarrow y''-\alpha_t$.
 \item Set $t\leftarrow t+1$.
\end{enumerate}
\end{enumerate}\end{small}

We first show that this procedure is well-defined. In every iteration, $\gamma>0$, and by induction $w^s\cdot x''=w^s\cdot y''=\gamma\cdot c_s$ for all $s\in[k]$. Thus in
step~\ref{step:fknap-1}, $LP_x$ (resp. $LP_y$) is non-empty: $x''/\gamma$ (resp. $y''/\gamma$) is a feasible solution. From the definition of $LP_x$ and $LP_y$ it also follows that $\delta>0$ in
step~\ref{step:fknap-2} and at least one coordinate of $x''$ or $y''$ is zeroed out in step~\ref{step:fknap-3}. This implies that the decomposition procedure terminates in $r\le 2n$ steps. At
the end of the procedure, we have decompositions $x'=\sum_{t=1}^r \beta_t$ and $y'=\sum_{t=1}^r \alpha_t$. Furthermore, each $\alpha_t$ (resp. $\beta_t$) corresponds to an {\em extreme point} of
$LP_y$ (resp. $LP_x$) in some iteration: hence the number of positive components in any of $\{\alpha_t,\beta_t\}_{t=1}^r$ is at most $k$. Finally note that for all $t\in [r]$, $w^s\cdot
\alpha_t=w^s\cdot \beta_t$ for all knapsacks $s\in[k]$.

Observe that Claim~\ref{cl:knap-local-opt} applies to $y$, $\alpha_t$ and $\beta_t$ (any $t\in [r]$) because each of $\alpha_t, \beta_t$ has support-size $k$, and $y-\alpha_t+\beta_t\in \us$ and
satisfies all knapsacks with equality. Strictly speaking, Claim~\ref{cl:knap-local-opt} requires the original local optimal $\tilde{y}$, which is not augmented with dummy elements. However even
$\tilde{y}-\alpha_t+\beta_t\in \us$ and satisfies all knapsacks (possibly not at equality), and the claim does apply. This gives:
\begin{equation}\label{eq:local-knap}
F(y-\alpha_t+\beta_t)\le (1+\epsilon)\cdot F(y) +
\frac{f_{max}}{4n^2} \qquad \forall t\in [r]~.
\end{equation}

Let $M\in\mathbb{Z}_+$ be large enough so that $M\alpha_t$ and $M\beta_t$ are integral for all $t\in[r]$. In the rest of the proof, we consider a scaled ground-set $U$ containing $M$ copies of
each element in $V$. We define function $g:2^U\rightarrow \mathbb{R}_+$ as $g(\cup_{i\in V} T_i) = F(\cdots,\frac{|T_i|}{M},\cdots)$ where each $T_i$ consists of copies of element $i\in V$.
Lemma~\ref{lem:scaled-submod} implies that $g$ is submodular. Corresponding to $y$ we have a set $P=\cup_{i\in V} P_i$ consisting of the first $|P_i|=M\cdot y_i$ copies of each element $i\in V$.
Similarly, $x$ corresponds to set $Q=\cup_{i\in V} Q_i$ consisting of the first $|Q_i|=M\cdot x_i$ copies of each element $i\in V$. Hence $P\cap Q$ (resp. $P\cup Q$) corresponds to $x\wedge y$
(resp. $x\vee y$) scaled by $M$. Again, $P\setminus Q$ (resp. $Q\setminus P$) corresponds to $y'$ (resp. $x'$) scaled by $M$. The decomposition of $y'$ from above suggests \emph{disjoint} sets
$\{A_t\}_{t=1}^r$ such that $\cup_t A_t =P\setminus Q$; i.e. each $A_t$ corresponds to $\alpha_t$ scaled by $M$. Similarly there are \emph{disjoint} sets $\{B_t\}_{t=1}^r$ such that $\cup_t B_t
=Q\setminus P$. Observe also that $g((P\setminus A_t)\cup B_t) = F(y-\alpha_t+\beta_t)$, so~\eqref{eq:local-knap} corresponds to:
\begin{equation}\label{eq:local-knap-scaled}
g((P\setminus A_t)\cup B_t)\le (1+\epsilon)\cdot g(P)+ \frac{f_{max}}{4n^2} \qquad \forall t\in [r] ~.
\end{equation}
Adding all these $r$ inequalities to $g(P)=g(P)$, we obtain $(r+\epsilon \cdot r+ 1)g(P) + \frac{r}{4n^2}f_{max}\ge g(P)+\sum_{t=1}^r g((P\setminus A_t)\cup B_t)$. Using submodularity of $g$ and
the disjointness of families $\{A_t\}_{t=1}^r$ and $\{B_t\}_{t=1}^r$, we obtain $(r+\epsilon \cdot r+ 1)\cdot g(P)+\frac{r}{4n^2}f_{max}\ge (r-1)\cdot g(P)+g(P\cup Q)+g(P\cap Q)$. Hence
$(2+\epsilon \cdot r)\cdot g(P)\ge g(P\cup Q)+g(P\cap Q)-\frac{r}{4n^2}f_{max}$. This implies the lemma because $r\le 2n$.
\end{pf}

\begin{theorem}\label{thm:fractional}
For any constant $\delta>0$, there exists a $({1\over 4}- \delta)$-approximation algorithm for problem~\eqref{prob:frac-k-knap} with all upper bounds $u_i=1$ (for all $i\in V$).
\end{theorem}

\noindent {\bf Rounding the fractional knapsack.} In order to solve the (non-fractional) submodular maximization subject to $k$ knapsack constraints, we partition the elements into two subsets.
For a constant parameter $\delta$, we say that element $e\in V$ is {\em heavy} if $w^s(e)\ge \delta$ for some knapsack $s\in[k]$. All other elements are called {\em light}. Since the number of
heavy elements in any feasible solution is bounded by $k\over \delta$, we can enumerate over all possible sets of heavy elements and obtain the optimal profit for these elements. For light
elements, we show a simple randomized rounding procedure (applied to the fractional solution from~\eqref{prob:frac-k-knap}) that gives a $({1\over 4}-\epsilon)$-approximation for the light
elements. Combining the enumeration method with the randomized rounding method, we get a $(\frac15-\epsilon)$-approximation for submodular maximization subject to $k$ knapsack constraints. The
details and proofs of this result are left to Appendix~\ref{app:knap-round}.

\section{Improved Bounds under Partition Matroids\label{sec:k-partn}}
The improved algorithm for partition matroids is again based on local search. In the \emph{exchange} local move of the general case (Section~\ref{sec:k-mat}), the algorithm only attempts to include one new element at a time
(while dropping upto $k$ elements). Here we generalize that step to allow including $p$ new elements while dropping up to $(k-1)\cdot p$ elements, for some fixed constant $p\ge 1$. We show that
this yields an improvement under partition matroid constraints. Given a current solution $S\in \cap_{j=1}^k \is_j$, the local moves we consider are:
\begin{itemize}
 \item {\bf Delete operation.} If $e\in S$ such that $f(S\setminus \{e\})>f(S)$, then $S\leftarrow
 S\setminus\{e\}$.
 \item  {\bf Exchange operation.} For some $q\le p$, if $d_1,\cdots,d_q\in V\setminus S$ and $e_i\in S\cup\{\phi\}$ (for $1\le i\le (k-1)\cdot q$)
 are such that: {\bf (i)} $S'=(S\setminus \{e_i ~:~ 1\le i\le (k-1)q\})\cup \{d_1,\cdots,d_q\}\in \is_j$ for all $j\in[k]$, and {\bf (ii)} $f(S')>f(S)$, then $S\leftarrow S'$.
\end{itemize}

\noindent The main idea here is a strengthening of Lemma~\ref{lem:k-mat-main}. Missing proofs of this section are in Appendix~\ref{app:k-partn}.
\begin{lemma}\label{lem:partn-mat}
 For a local optimal solution $S$ and any $C\in\cap_{j=1}^k \is_j$,
we have $k\cdot f(S)\ge (1-\frac1p)\cdot f(S\cup C)+(k-1)\cdot f(S\cap C)$.
\end{lemma}
\begin{pf}
We use an exchange property (see Schrijver~\cite{SchrijverBook}), which implies for any {\em partition matroid} $\m$ and $C,S\in\is(\m)$ the existence of a map $\pi:C\setminus S\rightarrow
(S\setminus C)\cup \{\phi\}$ such that
\begin{enumerate}
 \item $(S\setminus \{\pi(b) ~:~ b\in T\})\cup T \in\is(\m)$ for all $T\sse C\setminus S$.
 \item $|\pi^{-1}(e)|\le 1$ for all $e\in S\setminus C$.
\end{enumerate}
Let $\pi_j$ denote the mapping under partition matroid $\m_j$ (for $1\le j\le k$).

\noindent {\bf Combining partition matroids $\m_1$ and $\m_2$.} We use $\pi_1$ and $\pi_2$ to construct a multigraph $G$ on vertex set $C\setminus S$ and edge-set labeled by $E=\pi_1(C\setminus
S)\cap\pi_2(C\setminus S)\sse S\setminus C$ with an edge labeled $a\in E$ between $e,f\in C\setminus S$ iff $\pi_1(e)=\pi_2(f)=a$ or $\pi_2(e)=\pi_1(f)=a$. Each edge in $G$ has a unique label
because there is exactly one edge $(e,f)$ corresponding to any $a\in E$. Note that the maximum degree in $G$ is at most 2. Hence $G$ is a union of disjoint cycles and paths. We index elements of
$C\setminus S$ in such a way that elements along any path or cycle in $G$ are consecutive. For any $q\in \{0,\cdots,p-1\}$, let $R_q$ denote the elements of $C\setminus S$ having an index that
is \emph{not} $q$ modulo $p$. It is clear that the induced graph $G[R_q]$ for any $q\in [p]$ consists of disjoint paths/cycles, each of length at most $p$. Furthermore each element of
$C\setminus S$ appears in exactly $p-1$ sets among $\{R_q\}_{q=0}^{p-1}$.

\begin{cl}\label{cl:partn-mat}
For any $q\in \{0,\cdots,p-1\}$, $k\cdot f(S)\ge f(S\cup R_q)+(k-1)\cdot f(S\cap C)$.
\end{cl}

Adding the $p$ inequalities given by Claim~\ref{cl:partn-mat}, we get $pk\cdot f(S)\ge \sum_{q=0}^{p-1} f(S\cup R_q)+p(k-1)\cdot f(S\cap C)$. Note that each element of $C\setminus S$ is missing
in exactly $1$ set $\{S\cup R_q\}_{q=0}^{p-1}$, and elements of $S\cap C$ are missing in none of them. Hence an identical  simplification as in Lemma~\ref{lem:k-mat-main} gives $\sum_{q=0}^{p-1}
[f(S\cup C)-f(S\cup R_q)]\le f(S\cup C)-f(S)$. Thus,
$$(pk-1)\cdot f(S)\ge (p-1)\cdot f(S\cup C) +p(k-1)\cdot
f(S\cap C),$$ which implies $k\cdot f(S)\ge (1-\frac1p)\cdot f(S\cup C) + (k-1)\cdot f(S\cap C)$, giving the lemma.
\end{pf}

\begin{theorem}\label{thm:k-part}
For any $k\ge 2$ and fixed constant $\epsilon>0$, there exists a ${1\over k+1+\frac{1}{k-1}+\epsilon}$-approximation algorithm for maximizing a non-negative submodular function over $k$
partition matroids. This bound improves to ${1\over k+\epsilon}$ for monotone submodular functions.
\end{theorem}

\medskip
\noindent {\bf Acknowledgment:} The proof of Lemma~\ref{lem:k-mat-main} presented in this paper is due to Jan Vondr\'ak. Our original proof~\cite{lmns} was more complicated --- we thank Jan for
letting us present this simplified proof.

\begin{small}

\end{small}

\appendix

\section{Missing proofs from Section~\ref{sec:k-mat}}\label{app:k-mat}

\begin{pfof}{Theorem~\ref{th:mat-exch}}
We proceed by induction on $t=|J\setminus I|$. If $t=0$, there is nothing to prove. Suppose there is an element $b\in J\setminus I$ with $I\cup\{b\}\in{\cal I}({\cal M})$. In this case we apply
induction on $I$ and $J'=J\setminus \{b\}$ (where $|J'\setminus I|=t-1<t$). Because $I\setminus J'=I\setminus J$, we obtain a map $\pi':J'\setminus I\rightarrow (I\setminus J)\cup\{\phi\}$
satisfying the two conditions. The desired map $\pi$
%for $\langle I,J\rangle$
is then $\pi(b)=\phi$ and $\pi(b')=\pi'(b')$ for all $b'\in J\setminus I \setminus \{b\} = J'\setminus I$.

Now we may assume that $I$ is a maximal independent set in $I\cup J$. Let $\m'\sse\m$ denote the matroid $\m$ truncated to $I\cup J$; so $I$ is a base in $\m'$. We augment $J$ to some base
$\tilde{J}\supseteq J$ in $\m'$ (because any maximal independent set in $\m'$ is a base). Then we have two bases $I$ and $\tilde{J}$ in $\m'$. Theorem~39.12 from~\cite{SchrijverBook} implies the
existence of elements $b\in \tilde{J}\setminus I$ and $e\in I\setminus \tilde{J}$ such that both $(\tilde{J}\setminus b)\cup \{e\}$ and $(I\setminus e)\cup \{b\}$ are bases in $\m'$. Note that
$J'=(J\setminus \{b\})\cup \{e\}\sse (\tilde{J}\setminus \{b\})\cup \{e\}\in \m$. By induction on $I$ and $J'$ (because $|J'\setminus I|=t-1<t$) we obtain map $\pi':J'\setminus I\rightarrow I
\setminus J'$ satisfying the two conditions. The map $\pi$
%for $\langle I,J\rangle$
is then $\pi(b)=e$ and $\pi(b')=\pi'(b')$ for all $b'\in J\setminus I \setminus \{b\} = J'\setminus I$. The first condition on $\pi$ is satisfied by induction (for elements $J\setminus I
\setminus \{b\}$) and because $(I\setminus e)\cup \{b\}\in\m$ (see above). The second condition on $\pi$ is satisfied by induction and the fact that $e\not\in I\setminus J'$.
\end{pfof}

\begin{lemma}\label{lem:apx-local}
For an approximately  locally optimal solution $S$ (in procedure $B$) and any $C\in\cap_{j=1}^k \is_j$, $(1+\epsilon)(k+1)\cdot f(S)\ge f(S\cup C) + k\cdot f(S\cap C)$ where $\epsilon>0$ a
parameter defined in the algorithm description. Additionally for $k=1$, if $S\in\is_1$ is any locally optimal solution under only the swap operation, and $C\in \is_1$ with $|S|=|C|$, then
$2(1+\epsilon)\cdot f(S)\ge f(S\cup C) + f(S\cap C)$.
\end{lemma}
\begin{pf} The proof of this lemma is almost identical to the proof of the
Lemma \ref{lem:k-mat-main} the only difference is that left-hand sides of inequalities (\ref{eq:local1}) and inequalities (\ref{eq:local2}) are multiplied by $1+{\epsilon\over n^4}$.
Therefore, after following the steps in Lemma~\ref{lem:k-mat-main}, we obtain the inequality:
$$(k+1+{\epsilon\over n^4}\lambda)\cdot f(S)\ge f(S\cup C) + k\cdot f(S\cap C).$$

\noindent Since $\lambda\le (k+1)n$ (see Lemma~\ref{lem:k-mat-main}) and we may assume that $n^4>>(k+1)n$, we obtain the lemma.
\end{pf}

\noindent {\bf Running time of Algorithm $A$ (Theorem~\ref{th:main})} Here we describe a missing part of the proof of Theorem~\ref{th:main} about the running of Algorithm $A$. The parameter
$\epsilon>0$ in Procedure $B$ is any value such that $\frac1\epsilon$ is at most a polynomial in $n$. Note that using approximate local operations in the local search procedure B (in
Figure~\ref{fig:algorithmb}) makes the running time of the algorithm polynomial. The reason is as follows: one can easily show that for any ground set $X$ of elements, the value of the initial
set $S=\{v\}$ is at least $\opt(X)/ n$, where $\opt(X)$ is the optimal value of problem~\eqref{prob:k-mat} restricted to $X$. Each local operation in procedure $B$ increases the value of the
function by a factor $1+{\epsilon\over n^4}$. Therefore, the number of local operations for procedure $B$ is at most $\log_{1+{\epsilon\over n^4}} {\opt(X) \over {\opt(X) \over n}} = O({1\over
\epsilon}n^4\log n)$, and thus the running time of the whole procedure is ${1\over \epsilon} \cdot n^{O(k)}$. Moreover, the number of procedure calls of Algorithm $A$ for procedure $B$ is
polynomial, and thus the running time of Algorithm $A$ is also polynomial.

\section{Missing Proofs from Section~\ref{sec:k-knap}}\label{app:k-knap}

\begin{pfof}{Claim~\ref{cl:cont-prop}} Consider the line from $a_1$ to $a_2$ parameterized by
$t\in [0,1]$ as $y(t):= a_1+t (a_2-a_1)$. Observe that all points on this line are in $\dom$ (because $\dom$ is a convex set). At any $t\in [0,1]$, we have:
$$\frac{\partial l(y(t))}{\partial
t}=\sum_{j=1}^n \frac{\partial l(y(t))}{\partial x_j} \cdot \frac{\partial y_j(t)}{\partial t}=\sum_{j=1}^n\frac{\partial l(y(t))}{\partial x_j} \cdot (a_2(j)-a_1(j))\le 0.$$  Above, the first
equality follows from the chain rule because $l$ is differentiable, and the last inequality uses the fact that $a_2-a_1\ge 0$ coordinate-wise. This completes the proof of the claim.\end{pfof}

\begin{pfof}{Claim~\ref{cl:cont-submod}}
Let $\dom=\{y\in[0,1]^n ~:~ y+d\in [0,1]^n\}$. Define function $h:\dom\rightarrow \mathbb{R}_+$ as $h(x):= F(x+d)-F(x)$, which is a multi-linear polynomial. We will show that $\frac{\partial
h}{\partial x_i}(\alpha)\le 0$ for all $i\in V$, at every point $\alpha\in \dom$. This combined with Claim~\ref{cl:cont-prop} below would imply $h(a)\ge h(q)$ because $a\le q$ coordinate-wise,
which gives the claim.

In the following, fix an $i\in V$ and denote $F'_i(y)=\frac{\partial F}{\partial x_i}(y)$ for any $y\in [0,1]^n$. To show $\frac{\partial h}{\partial x_i}(\alpha)\le 0$ for $\alpha\in \dom$, it
suffices to have $F'_i(\alpha+d)-F'_i(\alpha)\le 0$. From the continuous submodularity of $F$, for every $j\in V$ we have $\frac{\partial F'_i}{\partial x_j}(y) = \frac{\partial^2 F}{\partial
x_j
\partial x_i}(y) \le 0$ for all $y\in [0,1]^n$. Then applying
Claim~\ref{cl:cont-prop} to $F'_i$ (a multi-linear polynomial) implies that $F'_i(\alpha+d)-F'_i(\alpha)\le 0$. This completes the proof of Claim~\ref{cl:cont-submod}.
\end{pfof}

\begin{pfof}{Lemma~\ref{lem:scaled-submod}}
To show submodularity of $g$, consider any two subsets $P=\cup_{i\in V} P_i$ and $Q=\cup_{i\in V} Q_i$ of $U$, where each $P_i$ (resp., $Q_i$) are copies of element $i\in V$. We have $P\cap
Q=\cup_{i\in V} (P_i\cap Q_i)$ and $P\cup Q=\cup_{i\in V} (P_i\cup Q_i)$. Define vectors $p,q,a,b\in [0,1]^n$ as follows:
$$p_i=\frac{|P_i|}{s_i},\quad q_i=\frac{|Q_i|}{s_i},\quad a_i=\frac{|P_i\cap
Q_i|}{s_i},\quad b_i=\frac{|P_i\cup Q_i|}{s_i} \qquad \forall i\in V.$$ It is clear that $p+q=a+b$ and $d:= p-a\ge 0$. Submodularity condition on $g$ at $P,Q$ requires $g(P)+g(Q)\ge g(P\cap
Q)+g(P\cup Q)$. But by the definition of $g$, this is equivalent to $F(a+d)-F(a)\ge F(q+d)-F(q)$, which is true by Claim~\ref{cl:cont-submod}. Thus we have established the lemma.
\end{pfof}

\begin{pfof}{Claim~\ref{cl:knap-local-opt}}
Let $z\le y'$ be the point in $\us\cap \gs^n$ that minimizes $\sum_{i=1}^n (y'_i-z_i)$. Note that $z$ is a feasible local move from $y$: it lies in $\gs^n$, satisfies all knapsacks and the
upper-bounds, and is obtainable from $y$ by reducing $k$ variables and increasing $k$ others. Hence by local optimality $F(z)\le (1+\epsilon)\cdot F(y)$.

By the choice of $z$, it follows that $|z_i-y'_i|\le \zeta$ for all $i\in V$. Suppose $B$ is an upper bound on all first partial derivatives of function $F$ on $\us$: i.e. $\left|\frac{\partial
F(x)}{\partial x_i}\left|_{\bar{x}}\right.\right|\le B$ for all $i\in V$ and $\bar{x}\in \us$. Then because $F$ has continuous derivatives, we obtain
$$|F(z)-F(y')|\le \sum_{i=1}^n B\cdot |z_i-y'_i|\le nB\zeta\le 2n^2f_{max}\cdot \zeta\le \frac{f_{max}}{4n^2}.$$
Above $f_{max}=\max\{f(v) ~:~ v\in V\}$. The last inequality uses $\zeta=\frac1{8n^4}$, and the second to last inequality uses $B\le 2n\cdot f_{max}$ which we show next. Consider any $\bar{x}\in
[0,1]^n$ and $i\in V$. We have
\begin{eqnarray*}
\left| \frac{\partial F(x)}{\partial x_i}  \left|_{\bar{x}}\right. \right| & ~=~ & \left|\sum_{S\sse [n]\setminus \{i\}} \left[f(S\cup \{i\}) - f(S)\right]\cdot \Pi_{a\in S} \bar{x}_a\cdot
\Pi_{b\in S^c\setminus i}
(1-\bar{x}_b)\right|\\
& ~\le~ & \max_{S\sse [n]\setminus \{i\}} \left[f(S\cup \{i\}) + f(S)\right] ~\le~ 2n\cdot f_{max}~.
\end{eqnarray*}
Thus we have $F(y')\le (1+\epsilon)\cdot F(y) + \frac1{4n^2} f_{max}$.
\end{pfof}

\begin{pfof}{Theorem~\ref{thm:fractional}}
Because each singleton solution $\{i\}$ is feasible for the knapsacks and upper bounds are $1$, we have a feasible solution of value $f_{max}$. Choose $\epsilon=\frac{\delta}{n^2}$. The
algorithm runs the fractional local search algorithm (with all upper bounds $1$) to get locally optimal solution $y_1\in [0,1]^n$. Then we run another fractional local search, this time with
each variable $i\in V$ having upper bound $u_i=1-y_1(i)$; let $y_2$ denote the local optimum obtained here. The algorithm outputs the better of the solutions $y_1$, $y_2$, and value $f_{max}$.

Let $x$ denote the globally optimal fractional solution to~\eqref{prob:frac-k-knap}, where upper bounds are $1$. We will show $(2+\delta)\cdot (F(y_1)+F(y_2))\ge F(x)-f_{max}/n$, which would
prove the theorem. Observe that $x'=x-(x\wedge y_1)$ is a feasible solution to the second local search. Lemma~\ref{lem:knap-local-sol} implies the following for the two local optima:
\begin{eqnarray*}
(2+\delta)\cdot F(y_1)& \ge & F(x\wedge y_1) + F(x\vee y_1)-\frac{f_{max}}{2n}~,\\
(2+\delta)\cdot F(y_2) & \ge & F(x'\wedge y_2) + F(x'\vee y_2)-\frac{f_{max}}{2n}~.
\end{eqnarray*}

We show that $F(x\wedge y_1) +F(x\vee y_1) +F(x'\vee y_2)\ge F(x)$, which suffices to prove the theorem. For this inequality, we again consider a scaled ground-set $U$ having $M$ copies of each
element in $V$ (where $M\in\mathbb{Z}_+$ is large enough so that $Mx$, $My_1$, $My_2$ are all integral). Define function  $g:2^U\rightarrow \mathbb{R}_+$ as $g(\cup_{i\in V} T_i) =
F(\cdots,\frac{|T_i|}{M},\cdots)$ where each $T_i$ consists of copies of element $i\in V$. Lemma~\ref{lem:scaled-submod} implies that $g$ is submodular. Also define the following subsets of $U$:
$A$ (representing $y_1$) consists of the first $My_1(i)$ copies of each element $i\in V$, $C$ (representing $x$) consists of the first $Mx(i)$ copies of each element $i\in V$, and $B$
(representing $y_2$) consists of $My_2(i)$ copies of each element $i\in V$ (namely the copies numbered $My_1(i)+1$ through $My_1(i)+My_2(i)$) so that $A\cap B=\phi$. Note that we can indeed pick
such sets because $y_1+y_2\le 1$ coordinate-wise. Also we have the following correspondences via scaling:
$$ A\cap C\equiv x\wedge y_1,\quad A\cup C\equiv x\vee y_1,\quad (C\setminus A)\cup B\equiv x'\vee
y_2~.$$

Thus it suffices to show $g(A\cap C)+g(A\cup C)+g((C\setminus A)\cup B)\ge g(C)$. But this follows from submodularity and non-negativity of $g$:
$$g(A\cap C)+g(A\cup C)+g((C\setminus A)\cup
B)\ge g(A\cap C)+ g(C\setminus A) + g(C\cup A\cup B) \ge g(C).$$
Hence we have the desired approximation for the fractional
problem~\eqref{prob:frac-k-knap}.
\end{pfof}

\section{Rounding the fractional solution under knapsack constraints}\label{app:knap-round}
Fix a constant $\eta>0$ and let $c=\frac{16}{\eta}$. We give a $(\frac15-\eta)$-approximation for submodular maximization over $k$ knapsack constraints, which is problem~\eqref{prob:k-knap}.
Define parameter $\delta=\frac{1}{4c^3k^4}$.  We call an element $e\in V$ {\em heavy} if $w^i(e)\ge \delta$ for some knapsack $i\in[k]$. All other elements are called {\em light}. Let $H$ and
$L$ denote the heavy and light elements in an optimal integral solution. Note that $|H|\le k/\delta$. Hence enumerating over all possible sets of heavy elements, we can obtain profit at least
$f(H)$ in $n^{O(k/\delta)}$ time, which is polynomial for fixed $k$. We now focus only on light elements and show how to obtain profit at least $\frac14\cdot f(L)$. Later we show how these can
be combined into an approximation algorithm for problem~\eqref{prob:k-knap}. Let $\opt\ge f(L)$ denote the optimal value of the knapsack constrained problem, restricted to only light elements.

{\bf Algorithm for light elements.} Restricted to light elements, the algorithm first solves the fractional relaxation~\eqref{prob:frac-k-knap} with all upper bounds $1$, to obtain solution $x$
with $F(x)\ge (\frac14-\frac\eta{2}) \cdot \opt$, as described in the previous subsection (see Theorem~\ref{thm:fractional}). Again by adding dummy light elements for each knapsack, we assume
that fractional solution $x$ satisfies all knapsacks with equality. Fix a parameter $\epsilon= \frac1{ck}$, and pick each element $e$ into solution $S$ independently with probability
$(1-\epsilon)\cdot x_e$. We declare failure if $S$ violates any knapsack and claim zero profit in this case (output the empty set as solution). Clearly this algorithm always outputs a feasible
solution. In the following we lower bound the expected profit. Let $\alpha(S):= \max\{ w^i(S) ~:~ i\in [k]\}$.

\begin{cl}\label{cl:knap-rnd1}
For any $a\ge 1$, $Pr[\alpha(S)\ge a] \le k\cdot e^{-cak^2}$.
\end{cl}
\begin{pf}
Fixing a knapsack $i\in[k]$, we will bound $Pr[w^i(S)\ge a]$. Let $X_e$ denote the binary random variable which is set to 1 iff $e\in S$, and let $Y_e=\frac{w^i(e)}{\delta} X_e$. Because we only
deal with light elements, each $Y_e$ is a $[0,1]$ random variable. Let $Z_i:= \sum_e Y_e$, then $E[Z_i]=\frac{1-\epsilon}{\delta}$. By scaling, it suffices to upper bound $Pr\left[Z_i\ge
a(1+\epsilon)E[Z_i]\right]$. Because the $Y_e$ are independent $[0,1]$ random variables, Chernoff bounds~\cite{MR-book} imply:
$$Pr\left[Z_i\ge a(1+\epsilon)E[Z_i]\right] \le e^{-E[Z_i]\cdot
a\epsilon^2/2 } \le e^{-a\epsilon^2/4\delta}= e^{-cak^2}.$$ Finally by a union bound, we obtain $Pr[\alpha(S)\ge a]\le \sum_{i=1}^k Pr[w^i(S)\ge a]\le k\cdot e^{-cak^2}$.
\end{pf}

\begin{cl}\label{cl:knap-rnd2}
For any $a\ge 1$, $\max\{f(S) ~:~ \alpha(S)\le a+1\}\le 2(1+\delta)k(a+1)\cdot \opt$.
\end{cl}
\begin{pf}
We will show that for any set $S$ with $\alpha(S)\le a+1$, $f(S)\le2(1+\delta)k(a+1)\cdot \opt$, which implies the claim. Consider partitioning set $S$ into a number of smaller parts each of
which satisfies all knapsacks as follows. As long as there are remaining elements in $S$, form a group by greedily adding $S$-elements until no more addition is possible, then continue to form
the next group. Except for the last group formed, every other group must have filled up some knapsack to extent $1-\delta$ (otherwise another light element can be added). Thus the number of
groups partitioning $S$ is at most $\frac{k(a+1)}{1-\delta}+1\le 2k(a+1)(1+\delta)$. Because each of these groups is a feasible solution, the claim follows by the subadditivity of $f$.\end{pf}

\begin{lemma}\label{lem:knap-rnd}
The algorithm for light elements obtains expected value at least $(\frac14-\eta)\cdot\opt$.
\end{lemma}
\begin{pf}
Define the following disjoint events: $A_0:= \{\alpha(S)\le 1\}$, and $A_l:= \{l< \alpha(S)\le 1+l\}$ for any $l\in \mathbb{N}$. Note that the expected value of the algorithm is $\alg=E[f(S)\mid
A_0]\cdot Pr[A_0]$. We can write:
$$F(x)=E[f(S)]=E[f(S)\mid A_0]\cdot Pr[A_0] + \sum_{l\ge 1} E[f(S)\mid A_l]\cdot
Pr[A_l]= \alg + \sum_{l\ge 1} E[f(S)\mid A_l]\cdot Pr[A_l].$$ For any $l\ge 1$, from Claim~\ref{cl:knap-rnd1} we have $Pr[A_l]\le Pr[\alpha(S)>l]\le k\cdot e^{-clk^2}$. From
Claim~\ref{cl:knap-rnd2} we have $E[f(S)\mid A_l]\le 2(1+\delta)k(l+1)\cdot \opt$. So,
$$E[f(S)\mid A_l]\cdot Pr[A_l]\le k\cdot e^{-clk^2}\cdot 2(1+\delta)k(l+1)\cdot
\opt\le 8\cdot \opt \cdot lk^2 \cdot e^{-clk^2}.$$ Consider the expression $\sum_{l\ge 1} lk^2\cdot e^{-clk^2}\le \sum_{t\ge 1}t\cdot e^{-ct}\le \frac{1}{c}$, for large enough constant $c$.
Thus:
$$
\alg= F(x) - \sum_{l\ge 1} E[f(S)\mid A_l]\cdot Pr[A_l] \ge F(x) - 8\cdot\opt \sum_{l\ge 1} lk\cdot e^{-clk}\ge F(x)-\frac8c \opt.
$$
Because $\eta=\frac{16}{c}$ and $F(x)\ge (\frac14-\frac\eta{2})\cdot \opt$ from Theorem~\ref{thm:fractional}, we obtain the lemma.\end{pf}

\begin{theorem}
For any constant $\eta>0$, there is a $(\frac15-\eta)$-approximation algorithm for maximizing a non-negative submodular function over $k$ knapsack constraints.
\end{theorem}
\begin{pf}
As mentioned in the beginning of this subsection, let $H$ and $L$ denote the heavy and light elements in an optimal integral solution. The enumeration algorithm for heavy elements produces a
solution of value at least $f(H)$. Lemma~\ref{lem:knap-rnd} implies that the rounding algorithm for light elements produces a solution of expected value at least $(\frac14-\eta)\cdot f(L)$. By
subadditivity, the optimal value $f(H\cup L)\le f(H)+f(L)$. The better of the two solutions (over heavy and light elements respectively) found by our algorithm has value:
$$\max\{f(H), (\frac14-\eta)\cdot f(L)\}\ge \frac15 \cdot f(H) + \frac45 \cdot(\frac14-\eta)\cdot
f(L) \ge (\frac15 -\eta) \cdot f(H\cup L).$$ This implies the desired approximation guarantee.\end{pf}

\section{Missing Proofs from Section~\ref{sec:k-partn}}\label{app:k-partn}

\begin{pfof}{Claim~\ref{cl:partn-mat}}
The following arguments hold for any $q\in [p]$, and for notational simplicity we denote $R=R_q\sse C\setminus S$. Let $\{D_l\}_{l=1}^t$ denote the vertices in connected components of $G[R]$,
which form a partition of $R$. As mentioned above, $|D_l|\le p$ for all $l\in [t]$. For any $l\in [t]$, let $E_l$ denote the labels of edges in $G$ incident to vertices $D_l$. Because
$\{D_l\}_{l=1}^t$ are distinct connected components in $G[R]$, $\{E_l\}_{l=1}^t$ are disjoint subsets of $E\sse S\setminus C$. Consider any $l\in [t]$: we claim $S_l=(S\setminus E_l)\cup
D_l\in\is_1\cap\is_2$. Note that $E_l\supseteq\{\pi_1(b) ~:~ b\in D_l\}$ and $E_l\supseteq\{\pi_2(b) ~:~ b\in D_l\}$. Hence $S_l\sse (S\setminus\{\pi_i(b) ~:~ b\in D_l\})\cup D_l$ for $i=1,2$.
But from the property of mapping $\pi_i$ (where $i=1,2$), $(S\setminus\{\pi_i(D_l))\cup D_l\in\is_i$. This proves that $S_l\in\is_1\cap\is_2$ for all $l\in [t]$.

From the properties of the maps $\pi_j$ for each partition matroid $\m_j$, we have $(S\setminus \pi_j(D_l))\cup D_l\in \is_j$ for each $3\le j\le k$. Thus the following sets are independent in
all matroids $\m_1,\cdots,\m_k$~:
$$\big(S\setminus (\cup_{j=3}^k \pi_j(D_l) \cup E_l)\big)\cup
D_l\qquad \forall~ l\in [t].$$

Additionally, because $|D_l|\le p$ and $|(\cup_{j=3}^k \pi_j(D_l) \cup E_l|\le (k-1)\cdot p$, each of the above sets are in the local neighborhood of $S$. But local optimality of $S$ implies:
\begin{equation}f(S) \ge f(\big(S\setminus (\cup_{j=3}^k \pi_j(D_l) \cup E_l)\big)\cup D_l)\qquad\qquad \forall l\in [t].\end{equation}
Recall that $\{E_l\}$ are disjoint subsets of $S\setminus C$. Also each element $i\in S\setminus C$ is missing in the right-hand side of $n_i\le k-1$ terms (the $\pi_j$s are `matchings' onto
$S\setminus C$). Using local optimality under deletions, we have the inequalities:
\begin{equation}\label{eq:partn-mat2}(k-1-n_i)\cdot f(S)\ge (k-1-n_i)\cdot f(S\setminus \{i\})\qquad \forall
i\in S\setminus C.
\end{equation}
Now, proceeding as in the simplification done in Lemma~\ref{lem:k-mat-main} (using disjointness of $\{D_l\}_{l=1}^t$), we obtain:
$$f\left(S\cup\left(\cup_{l=1}^t D_l\right)\right) - f(S) \le (k-1)\cdot \left( f(S) - f(S\cap C) \right)$$
Noting that $\cup_{l=1}^t D_l=R$, we have the claim. \ignore{Note that elements of $S\cap C$ are present in all of them. Adding these $t$ inequalities to $f(S)=f(S)$ and successively combining
terms $f(X), f(Y)$ using $f(X)+f(Y)\ge f(X\cup Y)+f(X\cap Y)$, we have
\begin{equation}\label{eq:partn-mat1}(t+1)\cdot f(S)\ge f(S) + \sum_{l=1}^t
f(\big(S\setminus (\cup_{j=3}^k \pi_j(D_l) \cup E_l)\big)\cup D_l) \ge f(S\cup R)+\sum_{l=1}^t f(S\setminus A_l).
\end{equation} where
$\{A_l\sse S\setminus C\}_{l=1}^t$ are such that each element $a\in S\setminus C$ appears in $n_a\le k-1$ sets from $\{A_l\}_{l=1}^t$ (using Fact~\ref{fact:set-opn}). Furthermore, elements of
$S\cap C$
 do not appear in any of $\{A_l\}_{l=1}^t$. The second inequality
also uses the fact that $\cup_{l=1}^t D_l=R$. Letting $\lambda=t+\sum_{a\in S\setminus C} (k-1-n_a)$, adding inequalities from~\eqref{eq:partn-mat1} and~\eqref{eq:partn-mat2}, we obtain
$$(\lambda +1)\cdot f(S)\ge f(S\cup R) + \sum_{i=1}^\lambda
f(S\setminus A_i).$$ where $\{A_i\}_{i=1}^\lambda$ are such that each element of $S\setminus C$ appears in exactly $k-1$ of them (and elements of $S\cap C$ appear in none of them). An identical
simplification procedure as Claim~\ref{cl:simpl-proc} of Lemma~\ref{lem:k-mat-main} gives the following which implies the claim.
$$
(\lambda+1)\cdot f(S)\ge f(S\cup R) + (\lambda-k+1)\cdot f(S) + (k-1)\cdot f(S\cap C).
$$}
\end{pfof}

\begin{pfof}{Theorem~\ref{thm:k-part}}
We set $p=1+\lceil \frac{2k}{\epsilon}\rceil$. The algorithm for the monotone case is just the local search procedure with $p$-exchanges. Lemma~\ref{lem:partn-mat} applied to local optimal $S$
and the global optimal $C$ implies $f(S)\ge (\frac1k - \frac1{pk})\cdot f(S\cup C)\ge (\frac1k - \frac1{pk})\cdot f(C)$ (by non-negativity and monotonicity). From the setting of $p$, $S$ is a
$k+\epsilon$ approximate solution.

For the non-monotone case, the algorithm repeats the $p$-exchange local search $k$ times as in Theorem~\ref{th:main}. If $C$ denotes a global optimum, an identical analysis yields
$\left(1+\frac1{p-1}\right)k^2\cdot f(S)\ge (k-1)\cdot f(C)$. This uses the inequalities $$\left(\frac{p}{p-1}\right)k\cdot f(S_i)\ge f(S_i\cup C_i)+(k-1)\cdot f(S_i\cap C_i)\qquad \forall 1\le i\le k,$$ where $S_i$
denotes the local optimal solution in iteration $i\in\{1,\cdots,k\}$ and $C_i=C\setminus \cup_{j=1}^{i-1} S_j$. Using the value of $p$, $S$ is a $\left(k+1+\frac{1}{k-1}+\epsilon\right)$-approximate
solution. Observe that the algorithm has running time $n^{O(k/\epsilon)}$.
\end{pfof}

We note that the result for monotone submodular functions is the first improvement over the greedy $1\over k+1$-approximation algorithm~\cite{NWF78II}, even for the special case of partition
matroids. It is easy to see that the greedy algorithm is a ${1\over k}$-approximation for \emph{modular} functions. But it is only a ${1\over k+1}$-approximation for monotone submodular
functions. The following example shows that this bound is tight for every $k\ge 1$. The submodular function $f$ is the coverage function defined on a family $\mathcal{F}$ of sets. Consider a
ground set $E=\{e ~:~ 0\le e\le p(k+1)+1\}$ of natural numbers (for $p\ge 2$ arbitrarily large); we define a family $\mathcal{F}=\{S_i ~:~ 0\le i\le k\}\cup \{T_1,T_2\}$ of $k+3$ subsets of $E$.
We have $S_0=\{e ~:~ 0\le e\le p\}$, $T_1=\{e ~:~ 0\le e\le p-1\}$, $T_2=\{p\}$, and for each $1\le i\le k$, $S_i=\{e ~:~ p\cdot i +1\le e\le p\cdot (i+1)\}$. For any subset $S\sse \mathcal{F}$,
$f(S)$ equals the number of elements in $E$ covered by $S$; $f$ is clearly monotone submodular. We now define $k$ partition matroids over $\mathcal{F}$: for $1\le j\le k$, the $j^{th}$ partition
has $\{S_0,S_j\}$ in one group and all other sets in singleton groups. In other words, the partition constraints require that for every $1\le j\le k$, at most one of $S_0$ and $S_j$ be chosen.
Observe that $\{S_i ~:~ 1\le i\le k\}\cup \{T_1,T_2\}$ is a feasible solution of value $|E|=p(k+1)+1$. However the greedy algorithm picks $S_0$ first (because it has maximum size), and gets only
value $p+1$.

\section{Matroid Base Constraints}\label{sec:mat-base}
A base in a matroid is any maximal independent set. In this section, we consider the problem of maximizing a non-negative submodular function over \emph{bases} of some matroid $\m$.
\begin{equation}
\max \left\{f(S) ~:~ S\in \bs(\m)  \right\}.
\end{equation}

We first consider the case of symmetric submodular functions.
\begin{theorem}
There is a $(\frac{1}{3}-\epsilon)$-approximation algorithm for maximizing a non-negative symmetric submodular function over bases of any matroid.
\end{theorem}
\begin{pf}
We use the natural local search algorithm based only on swap operations. The algorithm starts with any maximal independent set and performs improving \emph{swaps} until none is possible. From
the second statement of Lemma~\ref{lem:k-mat-main}, if $S$ is a local optimum and $C$ is the optimal base, we have $2\cdot f(S)\ge f(S\cup C) + f(S\cap C)$. Adding to this inequality, the fact
$f(S)=f(\overline{S})$ using symmetry, we obtain $3\cdot f(S)\ge f(S\cup C) + f(\overline{S})+f(S\cap C) \ge f(C\setminus S) + f(S\cap C)\ge f(C)$. Using an approximate local search procedure to
make the running time polynomial, we obtain the theorem.
\end{pf}

However, the approximation guarantee of this algorithm can be arbitrarily bad if the function $f$ is not symmetric. An example is the directed-cut function in a digraph with a vertex bipartition
$(U,V)$ with $|U|=|V|=n$, having $t\gg 1$ edges from each $U$-vertex to $V$ and $1$ edge from each $V$-vertex to $U$. The matroid in this example is just the uniform matroid with rank $n$. It is
clear that the optimal base is $U$; on the other hand $V$ is a local optimum under swaps.

We are not aware of a constant approximation for the problem of maximizing a submodular function subject to an arbitrary matroid base constraint. For a special class of matroids we obtain the
following.
\begin{theorem}\label{thm:mat-2base}
There is a $(\frac{1}{6}-\epsilon)$-approximation algorithm for maximizing any non-negative submodular function over bases of matroid $\m$, when $\m$ contains at least two disjoint bases.
\end{theorem}
\begin{pf}
Let $C$ denote the optimal base. The algorithm here first runs the local search algorithm using only swaps to obtain a base $S_1$ that satisfies $2\cdot f(S_1)\ge f(S_1\cup C) + f(S_1\cap C)$,
from Lemma~\ref{lem:k-mat-main}. Then the algorithm runs a local search on $V\setminus S_1$ using both exchanges and deletions to obtain an independent set $S_2\sse V\setminus S_1$ satisfying
$2\cdot f(S_2)\ge f(S_2\cup (C\setminus S_1)) + f(S_2\cap (C\setminus S_1))$. Consider the matroid $\m'$ obtained by contracting $S_2$ in $\m$. Our assumption implies that $\m'$ also has two
disjoint bases, say $B_1$ and $B_2$ (which can also be computed in polynomial time). Note that $S_2\cup B_1$ and $S_2\cup B_2$ are bases in the original matroid $\m$. The algorithm outputs
solution $S$ which is the better of the three bases: $S_1$, $S_2\cup B_1$ and $S_2\cup B_2$. We have
\begin{eqnarray*}
6f(S) & ~\ge~ &  2f(S_1)+2\left(f(S_2\cup B_1)+f(S_2\cup B_2)\right) ~\ge~ 2f(S_1) +2f(S_2)\\
& ~\ge~ & f(S_1\cup C) + f(S_1\cap C) +f(S_2\cup (C\setminus S_1)) ~\ge~ f(C).
\end{eqnarray*}
 The second inequality uses the
disjointness of $B_1$ and $B_2$.\end{pf}

A consequence of this result is the following.
\begin{corollary}
Given any non-negative submodular function $f:2^V\rightarrow \mathbb{R}_+$ and an integer $0\le c\le |V|$, there is a $(\frac{1}{6}-\epsilon)$-approximation algorithm for the problem $\max\{f(S)
~:~ S\sse V,~|S|=c\}$.
\end{corollary}
\begin{pf}
If $c\le |V|/2$ then the assumption in Theorem~\eqref{thm:mat-2base} holds for the rank $c$ uniform matroid, and the theorem follows. We show that $c\le |V|/2$ can be ensured without loss of
generality. Define function $g:2^V\rightarrow \mathbb{R}_+$ as $g(T)=f(V\setminus T)$ for all $T\sse V$. Because $f$ is non-negative and submodular, so is $g$. Furthermore, $\max\{f(S) ~:~ S\sse
V,~|S|=c\}=\max\{g(T) ~:~ T\sse V,~|T|=|V|-c\}$. Clearly one of $c$ and $|V|-c$ is at most $|V|/2$, and we can apply Theorem~\ref{thm:mat-2base} to the corresponding problem.
\end{pf}

\end{document}